# An optical micro-comb with a 50GHz free spectral range for photonic microwave true time delays

Xingyuan Xu,[1,8] Jiayang Wu,[1,8] Thach G. Nguyen,[2] Tania Moein,[1] Sai T. Chu,[3] Brent E. Little,[4] Roberto Morandotti,[5,6,7] Arnan Mitchell,[2] and David J. Moss[1*]

[1]*Centre for Micro-Photonics, Swinburne University of Technology, Hawthorn, VIC 3122, Australia*
[2]*ARC Centre of Excellence for Ultrahigh-bandwidth Devices for Optical Systems (CUDOS), RMIT University, Melbourne, VIC 3001, Australia*
[3]*Department of Physics and Material Science, City University of Hong Kong, Tat Chee Avenue, Hong Kong, China.*
[4]*State Key Laboratory of Transient Optics and Photonics, Xi'an Institute of Optics and Precision Mechanics, Chinese Academy of Science, Xi'an, China.*
[5]*INRS-Énergie, Matériaux et Télécommunications, 1650 Boulevard Lionel-Boulet, Varennes, Québec, J3X 1S2, Canada.*
[6]*National Research University of Information Technologies, Mechanics and Optics, St. Petersburg, Russia.*
[7]*Institute of Fundamental and Frontier Sciences, University of Electronic Science and Technology of China, Chengdu 610054, China.*
[8]*These authors contribute equally to this paper.*
*\*Corresponding author: dmoss@swin.edu.au*

We demonstrate significantly improved performance of a microwave true time delay line (TTDL) based on an integrated micro-ring resonator (MRR) Kerr optical comb source with a channel spacing of 49GHz, corresponding to 81 channels over the C-band. The broadband microcomb, with a record low free spectral range of 49GHz, results in a large number of comb lines for the TTDL, greatly reducing the size, cost, and complexity of the system. The large channel count results in a high angular resolution and wide beam steering tunable range of the phased array antenna (PAA). The enhancement of PAA's performance matches well with theory, corroborating the feasibility of our approach as a competitive solution towards implementing compact low-cost TTDL in radar and communications systems.

## INTRODUCTION

In modern radar and communications systems, photonic signal processing has attracted great interest due to its numerous intrinsic advantages such as broad operation bandwidth, low loss, and strong immunity to electromagnetic interference [1–5]. Microwave photonic true time delay lines (TTDLs), which can introduce multiple progressive time delays, are one of the basic building blocks in microwave photonic systems and have wide applications in phased array antennas (PAAs), microwave photonic filters, analog-to-digital or digital-to-analog conversion, as well as arbitrary waveform generation [6–21].

Extensive efforts have been devoted to developing diverse methods for TTDLs, including using: dispersive elements such as single-mode fibres (SMF) [22], dispersion compensation fibres [23], and fibre Bragg gratings (FBGs) [24–25]; slow-light devices based on stimulated Brillouin scattering (SBS) and integrated resonators [26–28]; wavelength conversion coupled with chromatic dispersion [29], and so forth [30–31]. Moreover, tunable progressive TTDLs based on a switch-controlled dispersive recirculating loop [32], fast sweeping lasers [33], and dispersion-tunable media [34] have also been investigated.

For PAAs, the number of radiating elements determines the beamwidth, and so, in order to enhance the angular resolution, a large channel number is required for the TTDL. Typically, discrete lasers arrays [26, 32–34] or FBG arrays [24–25] are employed for multiple TTDL channels, resulting in a significant increase in system cost and complexity. In turn, this greatly limits the number of channels in practical systems. Other schemes based on optical frequency combs (OFCs) [6] can mitigate this problem, but many approaches to generating OFCs, such as those based on cascaded electro-optical (EO) modulators [9,35–40] and Fabry-Perot EO modulators [41], for example, require external radio frequency (RF) sources, which still imposes considerable cost and complexity to the TTDL.

On the other hand, OFCs generated by high-Q micro-resonators, or micro-combs [42–47], offer a new generation of compact, low-cost, and highly efficient sources, thus bringing about huge possibilities to achieve high-performance TTDLs. Advantages of integrated micro-combs for TTDLs include the potential for a large number of channels (wavelengths), greatly reduced footprint and complexity, as well as significantly improved performance for delay-line structure-based microwave photonic systems.

Recently [48, 49] we demonstrated photonic microwave TTDLs based on an integrated micro-comb source with a free spectral range (FSR) of 200GHz, achieving high performance when applied to PAAs in terms of beam resolution and angular tuning range. In this paper, we significantly improve the performance of this device, both in terms of angular resolution and beamwidth, by employing a microcomb source with a 49GHz FSR, thus dramatically increasing the available number of channels from 21 to 81 over the C-band. In addition to improve the performance of the TTDL, this compact multi-wavelength source can significantly reduce the size, cost, and complexity of the device.

## OPERATION PRINCIPLE

Figure 1 shows a schematic of the microwave TTDL based on an integrated optical Kerr comb source. The TTDL contains two modules: the first generates the optical micro-comb using an integrated MRR while the second creates replicas of input RF signals at each wavelength followed by weighting and then wavelength dependent time delays (i.e., induced by the dispersive medium) using standard optical fibres to form the high-channel count TTDL for the subsequent phased array antenna.

The MRR used to generate the Kerr optical comb (Fig. 2(a)) was fabricated on a high-index doped silica glass platform using CMOS compatible fabrication processes [42]. First, high-index (n = ~1.7 at 1550 nm) doped silica glass films were deposited using plasma enhanced chemical vapour deposition (PECVD). Those were then patterned by deep UV photolithography and etched via reactive ion etching to form waveguides with exceptionally low surface roughness. Finally, silica (n = ~1.44 at 1550 nm) was deposited via PECVD as an upper cladding. Our device architecture used a vertical coupling scheme where the gap can be controlled via film growth — a more accurate approach than lithographic techniques [50]. The vertical gap (Fig. 2b) between the bus waveguide and the MRR was approximately 200nm. A scanning electron microscope image of the cross-section of the MRR before depositing the silica upper cladding is shown in Fig. 2(b). Other advantages of our platform for nonlinear optics include ultra-low linear loss (~0.06 dB·cm$^{-1}$), a relatively high nonlinear parameter (~233 W$^{-1}$·km$^{-1}$), and in particular negligible nonlinear loss up to extremely high intensities (~25 GW·cm$^{-2}$) [42–44]. The compact integrated MRR had a radius of ~592 μm with a free spectral range (FSR) of ~0.4 nm, i.e., ~49 GHz, as is indicated in Fig. 2(c). This is the smallest FSR device to date for which micro–combs have been used for microwave and RF applications. This FSR – a factor of 4 smaller than previous results [52, 53], enables up to four times the number of channels (as many as 81 channels) over the C-band. After packaging the input and output ports of the device with fibre pigtails, the total insertion loss was ~1 dB. Due to the ultra-low loss of our platform, the MRR features narrow resonance linewidths, as shown in Fig. 2(d), corresponding to a Q factor of ~1.2 million. Furthermore, in order to obtain optimal parametric gain, the MRR was designed to feature anomalous dispersion in the C-band [47].

In the first module, CW pump light from a tunable laser source was amplified by an erbium-doped fibre amplifier (EDFA) to pump the on-chip MRR. Before the MRR, a tunable optical bandpass filter and a polarization controller were employed to suppress the amplified spontaneous emission noise and adjust the polarization state, respectively. When the pump wavelength was tuned to one of the resonances of the MRR and the pump power was high enough to provide sufficient parametric gain, parametric oscillation in the MRR occurred, ultimately generating a Kerr optical comb with nearly equal line spacing [42–43]. In order to avoid resonance drift and maintain the wavelength alignment of the resonances to the pump light, the MRR was mounted on a temperature controlled stage.

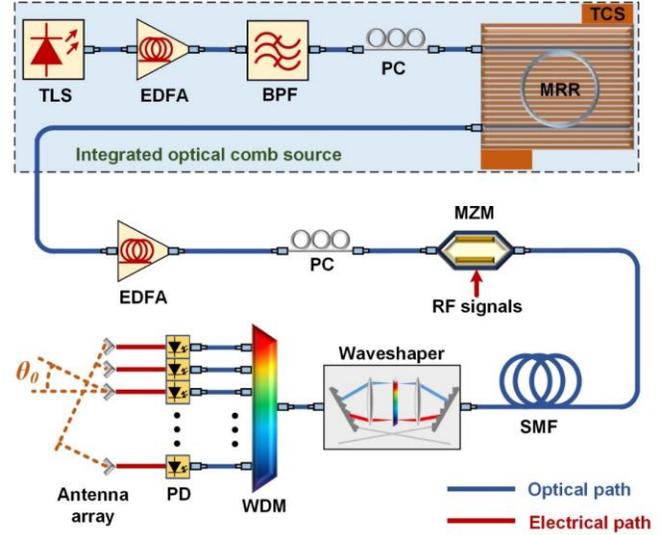

Fig. 1. Scheme of the proposed TTDL based on an integrated optical comb source. TLS: tunable laser source. EDFA: erbium-doped fibre amplifier. BPF: optical bandpass filter. PC: polarization controller. TCS: temperature controller stage. MZM: Mach-Zehnder modulator. SMF: single mode fibre. WDM: wavelength division multiplexer. PD: photodetector.

In the second module, the generated Kerr comb was directed to a Mach-Zehnder modulator (MZM), where replicas of an input RF signal were generated at each wavelength. The output optical signals from the MZM were delayed by a dispersive SMF (17 ps/nm/km), yielding a time delay difference between adjacent channels. Finally, individual channels were manipulated by a waveshaper, separated by a wavelength division multiplexer (WDM) and finally fed into the PAA system.

In PAA systems, optical signals on selected TTDL channels are separately converted into electrical domain and sent to an antenna array to form radiating elements. Considering that the antenna array is a uniformly spaced linear array with an element spacing of $d_{PAA}$, the steering angle $\theta_0$ of the PAA can be given as [52]

$$\theta_0 = \sin^{-1}\frac{c \cdot \tau}{d_{PAA}} \quad (1)$$

where $c$ is the speed of light in vacuum, and $\tau$ is the time delay difference between adjacent radiating elements. From Eq. (1), one can see that the steering angle can be tuned by adjusting $\tau$, i.e., changing the length of the dispersive medium or simply selecting every $m_{th}$ ($m$ = 1, 2, 3, …) TTDL channel as radiating elements. In the latter case, $\tau = mT$, where $T$ is the time delay difference between adjacent channels, which is determined jointly by the frequency spacing of the comb source and the dispersion accumulated from the dispersive medium. Thus, the steering angle is given by

$$\theta_0 = \sin^{-1}\frac{c \cdot mT}{d_{PAA}} \quad (2)$$

$$AF(\theta, \lambda_{RF}) = \frac{\sin^2[M\pi(d_{PAA}/\lambda_{RF})(\sin\theta - c\cdot mT/d_{PAA})]}{M^2\sin^2[\pi(d_{PAA}/\lambda_{RF})(\sin\theta - c\cdot mT/d_{PAA})]} \quad (3)$$

And the corresponding array factor (AF) of the PAA can be expressed as [52] where $\theta$ is the radiation angle, $M$ is the number of radiating elements, and $\lambda_{RF}$ is the wavelength of the RF signals. The angular resolution of the PAA is the minimum angular separation at which two equal targets at the same range can be separated, and is determined by the 3-dB beamwidth that can be approximated [53] as $\theta_{3dB} = 102/M$. This indicates that this quantity would greatly decrease with the number of radiating elements. Our TTDL based on an integrated optical comb source provides a large number of wavelengths for beam steering a large number of radiating elements, resulting in greatly enhanced angular resolution and tuning angle step resolution of the PAA. Compared with existing techniques based on discrete laser diode arrays, our approach relying on Kerr microcombs features a compact and simplified structure, with potentially significantly reduced cost, and with high performance resulting from the large number of delay channels.

## EXPERIMENTAL RESULTS

In order to generate a micro-comb with a 49 GHz-spacing the pump power was boosted up to 30.5 dBm via an EDFA and the wavelength was swept towards the red side. When the detuning between pump wavelength and MRR's resonance became small enough such that the intracavity field reached a threshold value, modulation instability was initiated, and primary combs arose with a spacing of multiple FSRs. As the detuning was further changed, the parametric gain lobes broadened and secondary comb lines with a spacing equal to the FSR of the MRR were generated by both degenerate and non-degenerate four wave mixing. This finally resulted a Type II Kerr optical comb [44, 54] (Fig. 2(e)) that was over 100-nm wide. A zoom-in view (green dashed) (Fig. 2(f)) indicates that potentially over 400 wavelengths were available for the TTDL. For our experiments, however, the wavelength range was limited to the C-band because of the waveshaper and EDFA, and so this resulted in 81 usable channels, or wavelengths. The generated comb served as the multi-wavelength source for the TTDL, which was working in a semi-coherent regime – we found that rigorous coherence of the comb was not crucial. Our comb envelope showed features that are typical of intermediate spectra, before the onset of cavity solitons [55, 56]. The key however is that the amplitude envelope was stable, and so the waveshaper could be used to flatten out the comb spectrum.

As Fig. 1 shows, the modulated signal after the MZM (EOSPACE) was propagated through ~2.122-km single mode (dispersive) fibre. The dispersion of the SMF was ~17.4 ps/(nm·km), corresponding to a time delay of ~14.8 ps between adjacent wavelength channels. Then we employed a Waveshaper (Finisar 4000s) as a wavelength division multiplexer to separate and weight these channels, and then convert back into the RF domain for PAA applications, thus achieving a high-channel count (up to 81 around the pump wavelength) TTDL with a compact structure.

The RF phase response was characterized by a vector network analyser (Anritsu 37369A), Fig. 3(a)), in which the channel at the central pump wavelength (channel 40) was set as the reference. The time delays corresponding to the measured phase slopes are shown in Fig. 3(b), in which a time delay step of ~14.8 ps/channel can be observed.

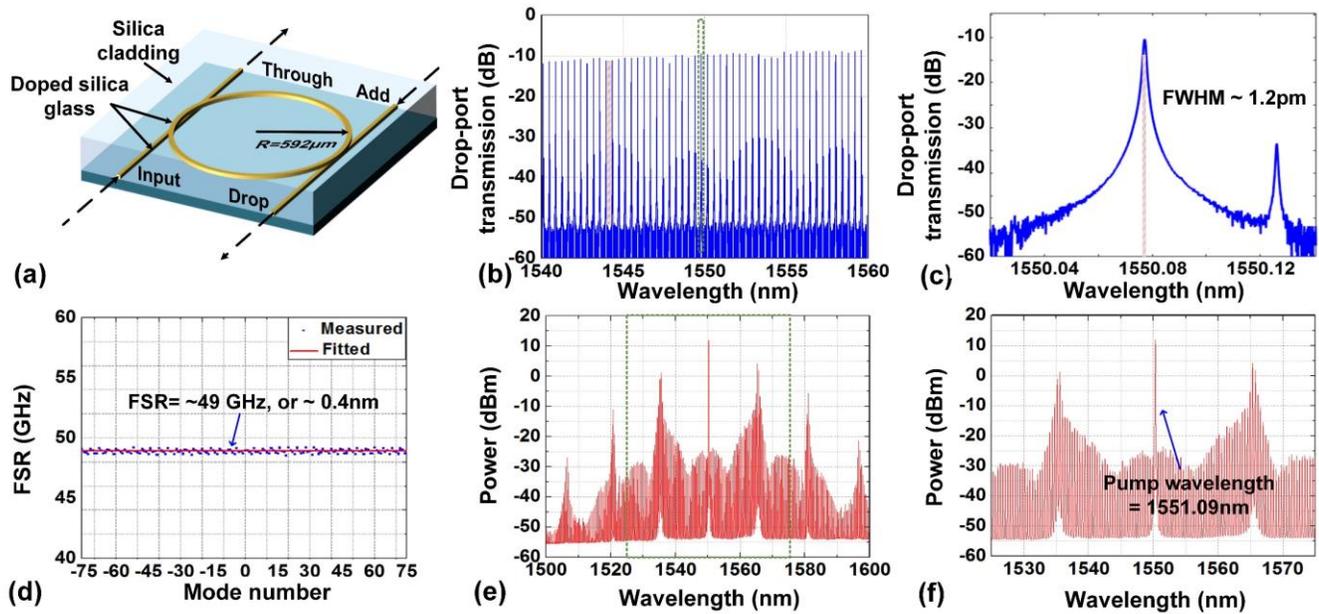

Fig. 2. (a) Schematic illustration of the MRR. Drop-port transmission spectrum of the on-chip MRR (b) with a span of 20 nm, showing an FSR of ~0.4 nm, and (c) a resonance at ~1550 nm with full width at half maximum (FWHM) of ~1.2 pm (~150 MHz). (d) Measured and fitted FSR of the MRR. Optical spectrum of the generated Kerr comb with a span of (e) 100nm and (f) 50 nm.

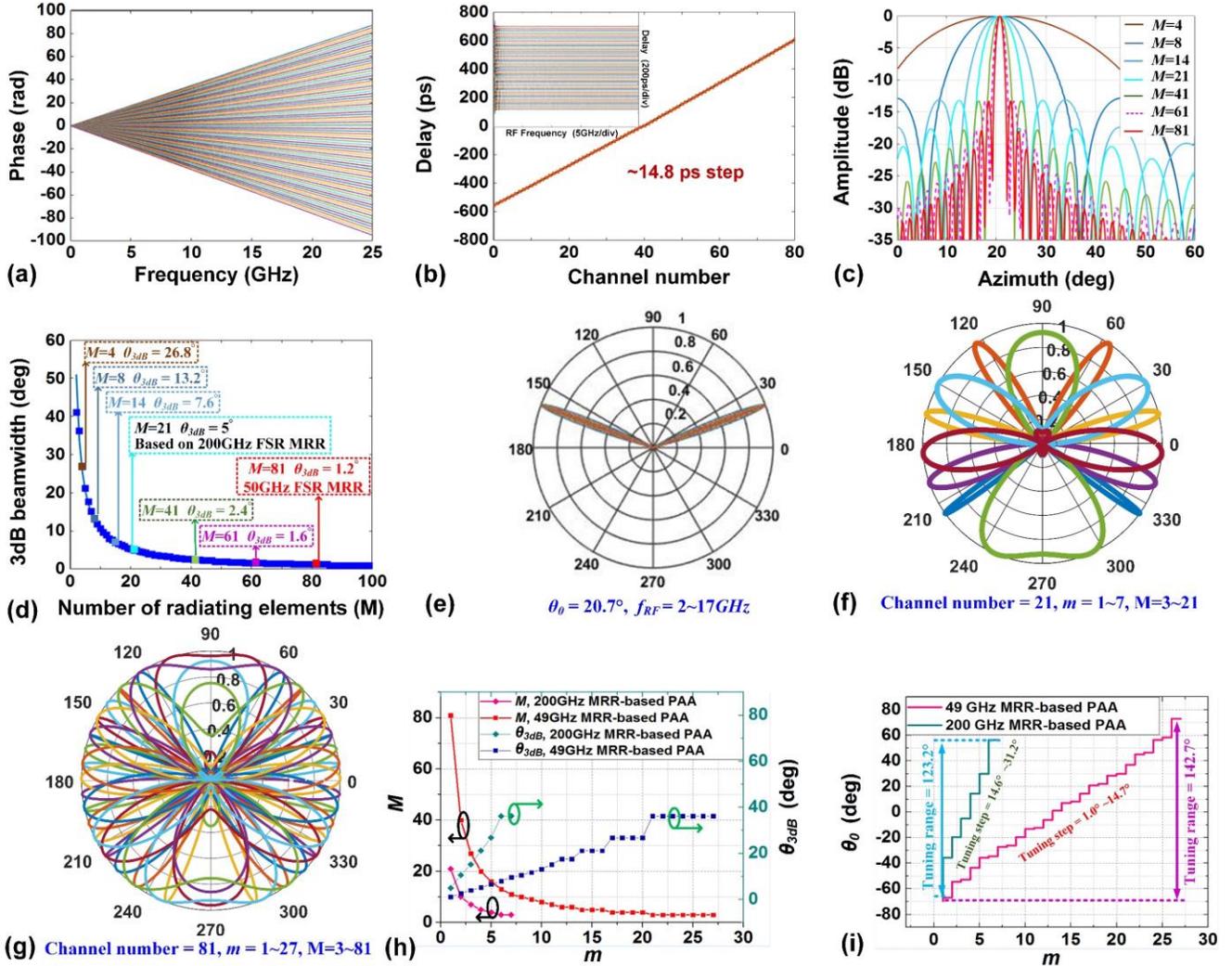

Fig. 3. (a) Measured RF phase responses of the 81-channel TTDL and (b) corresponding time delays of each channel; the inset shows flat delays over a wide RF frequency range. (c) Calculated array factors with $M$ varying from 4 to 81. (d) Relationship between number of radiating elements ($M$) and the 3dB beamwidth ($\theta_{3dB}$). (e) Calculated array factors with RF frequencies varying from 2 to 17GHz. (f) Calculated array factors of the PAA with $m$ varying from 1 to 7 based on a 200GHz-FSR Kerr comb [35]. (g) Calculated array factors of the PAA with $m$ varying from 1 to 27 based on the 49GHz-FSR Kerr comb. (h) Number of radiating elements ($M$) and the 3dB beamwidth ($\theta_{3dB}$) as a function of m. (i) The beam steering angle ($\theta_0$) as m varies.

To illustrate the enhancement in the PAA's performance, array factors (AFs) were calculated based on Eq. (3), where $\lambda_{RF}$ was set as the wavelength of the RF signal, $d_{PAA} = \lambda_{RF}/2$, and $T = 14.8$ ps. In Fig. 3(c), we note that, as $M$ increases from 4 to 81 (m=1), the corresponding 3dB beamwidth $\theta_{3dB}$ significantly decreases from 26.8° to 1.2°, matching well with the prediction $\theta_{3dB} = 102/M$, as shown in Fig. 3(d). Compared with previous results based on a 200GHz-FSR Kerr comb [48, 49], the much larger channel number provided by the 49GHz-FSR Kerr comb significantly improves the angular step size as well as beam resolution for the PAA. Moreover, the PAA can also achieve a wide instantaneous RF bandwidth without beam squint (i.e., variation in beam steering angle with RF frequency). Based on the proposed TTDL, as indicated in Fig. 3(e), the beam steering angle $\theta_0$ remains 20.7° while the RF frequency varies from 2 GHz to 17 GHz.

To achieve a tunable beam steering angle, every $m_{th}$ ($m = 1, 2, 3, ...$) wavelength of the TTDL could be selected by using the waveshaper. As well, the time delay ($\tau$) between the radiating elements could be varied with a step size of $T$. In Fig. 3(f–g), we present AFs of PAAs with varying $m$ and $M$, where $M$ = ⌊Channel Number / $m$⌋. Considering practical requirements in beam steering, $M$ can be set to 3 at least, thus for the 21-channel PAA based on the 200GHz-FSR Kerr comb [45], $m$ could reach 7 at most, while for the 81-channel PAA based on the 49 GHz-FSR Kerr comb, $m$ could reach as much as 27. Fig. 3. (h–i) shows the corresponding $M$, $\theta_{3dB}$ and $\theta_0$ as $m$ varies. As can be seen, on one hand, the 49GHz-FSR Kerr comb enables a much larger $M$ as $m$ varies, thus leading to a significantly smaller $\theta_{3dB}$ and greatly enhanced angular resolution (Fig. 3. (h)). On another hand, finer tuning steps (from 1.0° to 14.7°) as well as a large tuning range (142.7°) of the beam steering angle $\theta_0$ are now available due to the enhanced number of $m$ (Fig. 3.(i)).

It should be noted that the tunability of the beam steering angle could be further improved in terms of both tuning step and tuning range by employing dispersion tunable medium [32, 34] or more channels of the TTDL.

As a result, due to the large channel number of the proposed 49 GHz FSR microcomb-assisted TTDL, the PAA features greatly enhanced angular resolution, large instantaneous bandwidth, and a wide tunable range of the beam steering angle.

## CONCLUSION

In this paper, we demonstrate a high channel (81 over the C-band) TTDL based on an integrated Kerr comb source. A broadband Kerr comb with a large number of comb lines was generated by an on-chip MRR with an FSR of 49 GHz, and employed as a high-quality multi-wavelength source for the TTDL. Compared to traditional approaches, the size, complexity and, ultimately, the cost of the system can be greatly reduced. The large channel number of the TTDL resulted in the PAA featuring high angular resolution and wide range tunability of the beam steering angle. The enhancement in performance matches well with theory, confirming the feasibility of our approach as a promising solution towards implementing highly reconfigurable TTDLs for microwave photonic signal processing functions.

## ACKNOWLEDGMENTS

This work was supported by the Australian Research Council Discovery Projects Program (No. DP150104327). RM acknowledges support by the Natural Sciences and Engineering Research Council of Canada (NSERC) through the Strategic, Discovery and Acceleration Grants Schemes, by the MESI PSR-SIIRI Initiative in Quebec, and by the Canada Research Chair Program. He also acknowledges additional support by the Government of the Russian Federation through the ITMO Fellowship and Professorship Program (grant 074-U 01) and by the 1000 Talents Sichuan Program in China. Brent E. Little was supported by the Strategic Priority Research Program of the Chinese Academy of Sciences, Grant No. XDB24030000.